\documentclass[reprint,aps,prl]{revtex4-1}
\pdfoutput=1

\usepackage{amsmath}
\usepackage{amssymb}
\usepackage{graphicx}

\newcommand{\hk}{\text{HK} }
\newcommand{\hksup}{\text{HK}}
\newcommand{\la}{\left\langle}
\newcommand{\ra}{\right\rangle}
\newcommand{\as}{{\rm asym}}
\newcommand{\SU}{\text{SU}}
\newcommand{\R}{\mathbb{R}}
\DeclareMathOperator{\Tr}{Tr}

\begin{document}

\preprint{}
\title{Non-analyticity in scale in the planar limit of QCD}

\author{R.~Lohmayer}
\email{lohmayer@physics.rutgers.edu}
\author{H.~Neuberger}
\email{neuberg@physics.rutgers.edu}
\affiliation{Rutgers University, Department of Physics and Astronomy,
Piscataway, NJ 08854, USA}

\begin{abstract}
Using methods of numerical Lattice Gauge Theory we
show that in the limit of a large number of colors, 
properly regularized Wilson loops have an eigenvalue distribution which 
changes non-analytically as the overall size of the loop is
increased. This establishes a large-$N$ phase transition in continuum
planar gauge theory, a fact whose precise implications remain to be
worked out. 
\end{abstract}


\maketitle

Intuitively, one expects parallel transport round a closed curve in four-dimensional $\SU(N)$
pure gauge theory with $\theta_{CP} =0$  
to be close to identity for small curves and far from
identity for large curves. In this letter 
we make this idea concrete and find 
that small and large loops are separated by a large-$N$ 
phase transition. The possibility of a new type of non-analyticity entering 
$\SU(N)$ gauge theory in the 't Hooft limit~\cite{thooft} $N\to\infty$ has preoccupied researchers
for a long time; here we present a class of examples where this phenomenon occurs.
The numerical evidence is sufficiently convincing to view the effect
as an exact property of the large-$N$ limit.
We start presenting a simple lattice result 
and let it lead us to the above conclusion. 
After that, we put the result in historical context, 
mentioning some of the main papers this work is related 
to. 

\begin{figure}[hbt]
\begin{center}
\includegraphics[width=0.43\textwidth]{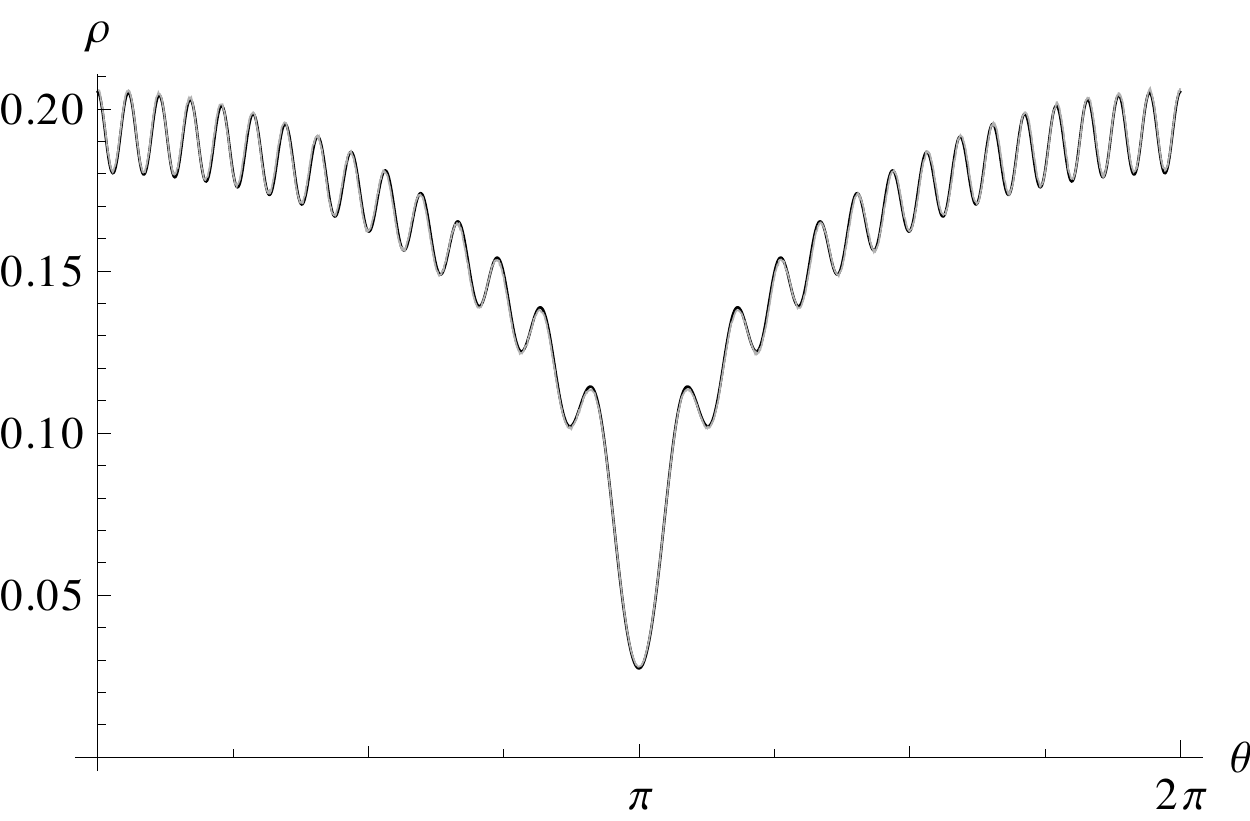}
\caption{Histogram of eigenvalue angles for a Wilson loop.}
\label{fig-histogram}
\end{center}
\end{figure}

The gray line in Fig.~\ref{fig-histogram} is the histogram for 
the angles of the eigenvalues of a square loop in $\SU(29)$ lattice gauge theory with side 
length equivalent to about $0.54$ fermi in QCD units and viewed at a resolution 
(thickness) of about $0.15$ fermi.
The black line is the eigenvalue-angle density determined by the heat-kernel
function (\hksup) for $\SU(29)$ which depends on a single  parameter $t$, to
be defined later (in the plot, $t\approx 4$). The two curves 
cannot be distinguished in the figure, so the  \hk approximates the data well, after adjusting $t$. The data is sufficiently accurate to check whether the \hk 
might provide an exact description. This is definitely ruled out by a $\chi^2$ analysis. 

Some definitions are in order now. The 
Wilson loop matrix associated with a closed spacetime curve $\mathcal
  C$ is
\begin{align*}
W_r(\mathcal C,x,s)=\mathcal{P} \exp\left(i\oint^x_{x;\mathcal{C}} A^r_\mu(y,s) dy_\mu
\right) \in \SU(N),
\end{align*}
where $r$ denotes an $\SU(N)$ representation, $x$ is a point on $\mathcal{C}$, and 
$s>0$ denotes a ``smearing parameter'' of dimension length squared ($\sqrt{s}$ determines the thickness 
of the loop). Smearing
is required to make all $W_r(\mathcal{C},x,s)$ finite $\SU(N)$ matrices 
with operator-valued entries. 
The gray line on the previous plot shows 
\begin{align*}
\rho_{N}(\theta;\mathcal{C},s)=\frac{1}{2\pi N} \sum_{i=1}^N \la \delta(\theta-\theta_i(\mathcal{C},s)\ra .
\end{align*}
The $\theta_i(\mathcal{C},s)$ are angles locating the eigenvalues of $W_f(\mathcal{C},x,s)$ 
($f$ denotes the fundamental representation) on the unit circle; they do not
depend on the choice of $x$. After averaging, also the dependence on the location and the orientation 
of the loop drops out. 
With the CP violating $\theta_{CP}$ parameter set to 0, $\rho_N$ is invariant under
$\theta\to 2\pi-\theta$. 

Smearing is defined as follows: 
One starts with five-dimensional gauge fields 
on $\R^4\times \R_+$; the smearing parameter $s$ lives on the $\R_+$.
The usual quantum fields are denoted by $B^f_\mu(x)$ and reside on the $\R^4$-boundary. The $A^f_\mu(x,s)$ are 
defined for $s\ge 0$ by
\begin{align*}
F^f_{\mu,s} = D^{\text{adjoint}}_{\nu} F^f_{\mu,\nu} \quad{\rm with}\quad A^f_\mu(x,s=0)=B^f_\mu(x).
\end{align*}
The 5D gauge freedom is reduced to a 4D one by $A^f_s(x,s)=0$. 
At $s>0$, all divergences coming from coinciding spacetime 
points in products of renormalized elementary fields
are eliminated by a limitation on the resolution 
of the observer, parametrized by $s$. Renormalization 
of the boundary quantum-fields $B^f_\mu(x)$ proceeds as usual. 
The definition of smearing easily extends to any finite UV 
cutoff including the lattice: replace $D_{\nu} F_{\mu,\nu}$ by 
the variation of the action with the UV cutoff in place. 
Smearing extends formally to loop space, with $\mathcal{C}$ parametrized by
$\sigma$,
\begin{align*}
\partial_s\! \Tr\!\la W_f(\mathcal{C},s)\ra=\oint_\sigma\! 
\frac{\delta^2 \Tr\!\la W_f(\mathcal{C},s)\ra}{\delta x_\mu^2 (\sigma)}\equiv
     {\hat L} \Tr\! \la W_f(\mathcal{C},s)\ra.
\end{align*}
${\hat L}$ is the L\'{e}vy Loop Laplacian appearing in the
Makeenko-Migdal (MM) equations.
Were a string representation of ${\hat L}$ found, 
diffusion in loop space would become well defined and field theory
and string theory could refer then to the same non-singular object. 

The \hk (heat kernel) probability density (w.r.t.~the Haar measure) for an $\SU(N)$ matrix $W$ is
\begin{align*}
\mathcal{P}_N^\hksup(W,t)&=\sum_{\text{all irred.}\ r} d_r \chi_r(W) e^{-\frac tN C_2(r)},
\end{align*}
implying $\la \chi_r(W) \ra = d_r e^{-\frac tN C_2(r)}$. 
The parameter $t$ is a ``diffusion time'' and $d_r$, $C_2(r)$ 
are the dimension and the quadratic Casimir of the irreducible representation $r$ (in Fig.~\ref{fig-histogram}, $t=3.881$). The heat kernel represents a multiplicative random walk on 
the $\SU(N)$ group manifold emanating from the identity.
The \hk single eigenvalue distribution $2\pi \rho_N^{\hksup}(\theta, t)$ is
given by
\begin{align*}
 1+\frac{2}{N}\sum_{p=0}^{N-1} (-1)^p
\sum_{q=0}^\infty \cos((p+q+1)\theta)d(p,q) e^{-\frac{t}{N} C(p,q)} ,
\end{align*}
where $C(p,q)=\frac{1}{2}(p+q+1)\left ( N -\frac{p+q+1}{N} + q-p\right )$ and 
$d(p,q)=\frac{(N+q)!}{p!q!(N-p-1)!}\frac{1}{p+q+1}$.

The \hk represents the data very well, but is not exact. One
could explain this by postulating ``Casimir dominance'':
$\Tr\la W_r(\mathcal C,s)\ra\approx  d_r \; e^{-C_2(r) {\mathfrak S}(\mathcal C,s)}$, with a
$r$-independent $\mathfrak S (\mathcal C,s)$. 
This approximation must break down for very large loops,
where screening effects come in and the loop is dominated
by the area term. Then, only the $N$-ality of $r$ should matter.
However, both in perturbation theory and at intermediate scales (up to 2 fermi),
Casimir dominance is known to be a good approximation. All our data is in the range (0,1) fermi because 
the large-$N$ transition is roughly in the middle of this segment. 

We proceed to make the case that this indication of approximate Casimir
dominance can be replaced by an exact statement in the large-$N$ limit, namely that 
the \hk formula is exact in a precise ``large-$N$ universality'' sense to be
explained below. The main point is that the shape of the eigenvalue-angle 
distribution of smeared Wilson matrices associated with uncomplicated loops
is governed by two opposite tendencies: random-matrix eigenvalue repulsion
and asymptotic-freedom attraction to unity. This produces $N$ alternating 
peaks and valleys with a swing of order $\frac{1}{N}$ between them. So, most of
the structure one sees in Fig.~\ref{fig-histogram} is determined by a generic mechanism. 
Rather than the oscillations, the truly 
interesting feature is the deep well around $\pi$. For a smaller loop, 
this well would be deeper and wider. Taking $N$ to infinity eliminates the 
more obvious details and leaves the essential features: for a small
loop the valley around $\pi$ flattens out at $N=\infty$ and the eigenvalue density is
zero there. For a large loop, the eigenvalue density is non zero around the
entire unit circle. In the \hk case, the role of overall loop
size is played by $t$ and the transition between a gap-less and a gapped spectrum 
occurs at $t=4$. Large-$N$ universality
will be a statement about the large-$N$ behavior of the gauge theory data in the neighborhood of a
loop size that would be critical at $N=\infty$. 

To concentrate on the region of interest we need to choose an observable 
that is particularly sensitive to eigenvalues close to -1:
\begin{align*}
\mathcal{O}_N(y,\mathcal C,s)=\la \det\left(e^{\frac y2}+e^{-\frac y 2 }
W_f(\mathcal C,s)\right)\ra .
\end{align*}
This observable generates all Wilson loop expectation values in the $k$-antisymmetric
irred.~representations of $\SU(N)$ and is given by
$\sum_{k=0}^N e^{\left(\frac N2 -k\right)y}\la\chi_k^\as(W_f (\mathcal C,s))\ra$. 

\begin{figure}[htb]
\begin{center}
    \includegraphics[width=0.43\textwidth]{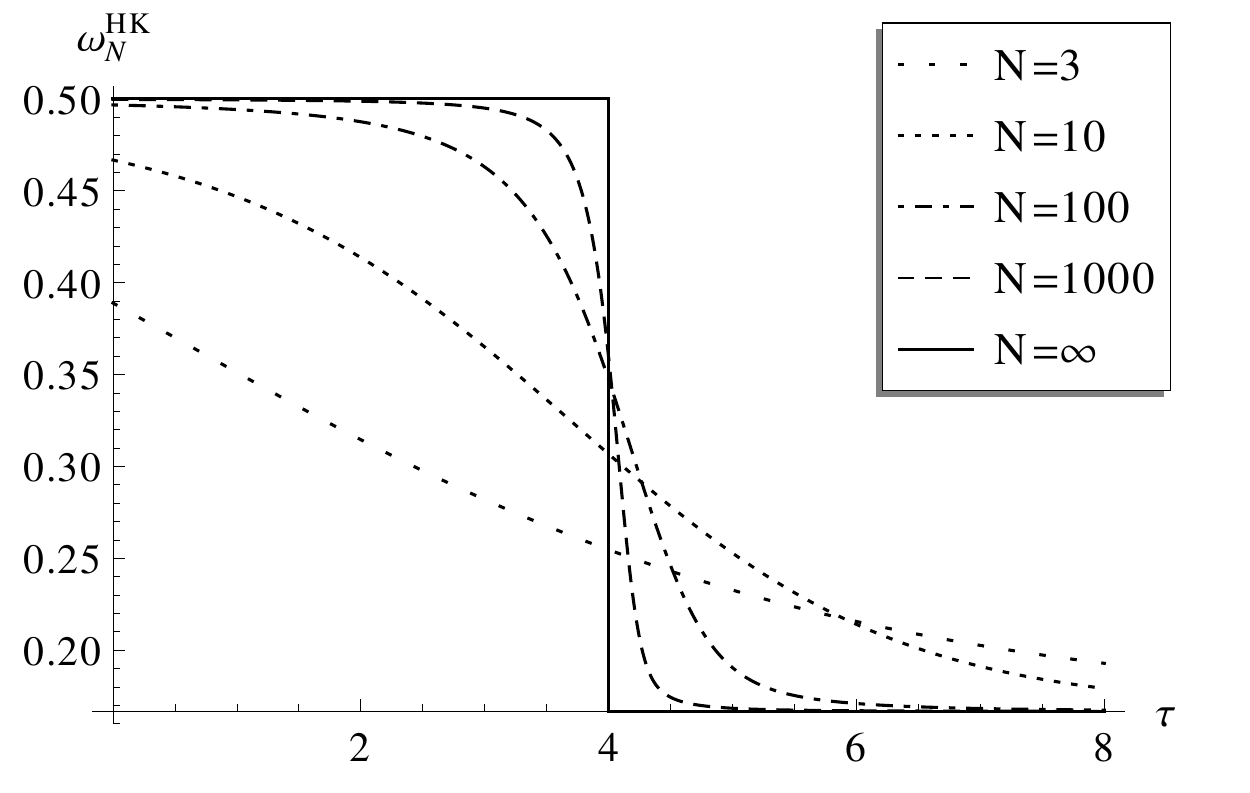}   
\caption{Development of a jump singularity in the \hk case.}
\label{fig-HKomegatau}
\end{center} 
 \end{figure}

At the transition point, $\log{\mathcal O}_N$ 
will have a non-analyticity at ${y=0}$ when $N=\infty$.  Therefore, we consider the expansion
\begin{align*}
\mathcal{O}_N(y,\mathcal C,s) = a_0(\mathcal C,s)+ a_1(\mathcal C,s) y^2 + a_2(\mathcal C,s) y^4 +O\left(y^6\right)
\end{align*}
and use 
$\omega_N(\mathcal C,s) = \frac{a_0(\mathcal C,s) a_2(\mathcal
  C,s)}{a_1^2(\mathcal C,s)}$ as our new observable. 
In the \hk case, with $\tau\equiv t (1+1/N)$, 
$\phi_N^\hksup(y,\tau)=-\frac{1}{N} \left ( \frac\partial{\partial y} \log(\mathcal{O}_N^\hksup(y,\mathcal \tau)
\right )$ obeys Burgers' equation:
\begin{align*}
\partial_\tau \phi_N^\hksup+\phi_N^\hksup\partial_y\phi_N^\hksup=\frac1{2N}\partial_y^2 \phi_N^\hksup.
\end{align*} 
$N$ enters 
only via the viscosity, which is $\frac{1}{2N}$. We may replace
the term ``large-$N$ universality'' by ``Burgers universality''. 
Our observable in the \hk case is plotted in Fig.~\ref{fig-HKomegatau} for different values of $N$.

Figure~\ref{fig-HKomegatau} shows that  $\omega_N^\hksup(\tau)$ becomes a $\theta$-function
at infinite $N$ but the singular behavior develops slowly with $N$. 
Burgers' equation implies that the jump in the $\theta$-function causes 
$\partial_y \phi_\infty^\hksup (y=0,\tau)$, 
which is finite and negative for $\tau <4$, to blow
up when $\tau$ reaches $\tau=4$. 
We expect a similar behavior in the four-dimensional gauge theory. To be specific, 
Burgers universality means, e.g., that the following
predictions about the large-$N$ limit hold exactly in the four-dimensional gauge theory, where criticality 
sets in at an overall loop size $l=l_c$: 
$\lim_{N\to\infty}{ N^{-\frac 32}} \left. \frac{a_1}{a_0}\right|_{l=l_c}=\frac 18 \sqrt{\frac
  32} \frac 1K$, (here, $~K\equiv\frac{1}{4\pi} \Gamma^2\left( \frac{1}{4}
\right ) \approx 1.046$); 
$\lim_{N\to\infty} {N^{-\frac32}} \left.\frac{a_2}{a_1}\right |_{l=l_c} = 
\frac{1}{24}\sqrt{\frac{3}{2}}K$;
$\lim_{N\to\infty}  \left. \omega_N \right |_{l=l_c} =\frac{1}{3}
K^2$. Here, the variation w.r.t.~the overall size $l$ is taken at constant loop shape. 
The roots of $\mathcal O_N(y)$ are all on the 
imaginary axis (as a consequence of the Lee-Yang theorem). Another universal property is that  
in the critical regime (around $y=0$,
$l=l_c$) these roots scale like ${N^{-\frac 34}}$.

Our objective was to establish numerically that the transition occurs and that
the universal predictions hold in the continuum limit of a lattice
gauge theory with standard single plaquette Wilson action. The lattice coupling,
traditionally denoted by $\beta$, is determined by the inverse 't Hooft coupling
$b=\frac\beta{2N^2}$. As $N$ varies, simulations are carried out in varying
ranges of $b$, all contained in the segment $[0.348,0.380]$ with upper limits determined by the lattice volumes, $V$.
We have carried out simulations at $N=11$, $19$, $29$ on 
hypercubic lattice volumes, $V$, $(12^4,14^4,18^4)$, $(10^4,12^4,13^4,14^4)$,
$(8^4,10^4,12^4)$, respectively. The volume size
determined the maximal $b$ allowed in order 
to stay in the confined phase at infinite $N$. For finite and fixed $N$, at $b$ values close to 
the maximal allowed one, sizable finite-volume effects became evident; they were consistent
with an exponential dependence on the linear size of the system. 
The data at different volumes was thus used to eliminate results
contaminated by finite-size effects. 

The measured observables were extracted from square $L\times L$ Wilson 
loops, with $L=1$, $2,\ldots$, $9$. Smearing was implemented on the lattice with a
parameter $S=L^2/55$ (our convention is 
that capital letter symbols correspond to
lower case symbols in the continuum). For a given Wilson loop, we
extracted the coefficients $a_{0,1,2}$ for each 
orientation and location of the loop. After averaging we obtained one set of 
coefficients for each gauge
configuration. We generated 160 independent gauge configurations at each set of 
parameters $(b,N,V)$. Consecutive $b$-values were spaced by increments $\Delta b=0.001$ and
were separated by 1000 passes, half of heat-bath type and half of over-relaxation type. 
Looking at autocorrelations, we determined that they typically drop by $e$ for every 250
such passes. Averaging over configurations 
produced coefficients which determined our estimate for $\omega_N(b,L)$,
and their errors. The data were sufficiently closely spaced in $b$ that we could use spline interpolation to 
represent the result as a continuous function of $b$. 

From the \hk case we learned that prohibitively large values of $N$ would
be needed to display directly the singular large-$N$ behavior. We adopted 
instead the following strategy: From each $\omega_N(b,L)$ we constructed 
a number $\tau_N (b,L)$ by solving $\omega_N(b,L)=\omega_{N}^\hksup
(\tau_N(b,L))$. The required inversion is unique. 
We then showed numerically that the convergence of $\tau_N(b,L)$ to
$\tau_\infty (b,L)$ is rapid, like that of the string tension 
and the deconfinement temperature. 
Hence, $\omega_\infty (b,L)=\lim_{N\to\infty}\omega_N^\hksup (\tau_N(b,L))=
\lim_{N\to\infty}\omega_{N}^\hksup (\tau_\infty(b,L))$ and 
to some universal subleading order in $\frac 1N$,
$\omega_N(b,L)\approx \omega_{N}^\hksup (\tau_\infty(b,L))$. 
This relation can be taken over to the continuum. 
Extrapolating to the continuum limit, 
the two variables $b$ and $L$ get
replaced by a single length variable, $l$, the side of the loop in 
physical units. We took measurements in a region in which the emerging
$\tau_\infty(l)$ extends on both sides of $\tau_\infty=4$. We numerically
determined that $\tau_\infty(l)$ is smooth in $l$ and invertible at $l=l_c$ where
$\tau_\infty (l_c)=4$. We conclude that the continuum $\omega_N(l)$
will develop the same singularity in the vicinity of $l=l_c$ as 
$\omega_N^\hksup (\tau_\infty (l))$ would, establishing the transition and its
universality. 

\begin{figure}[htb]
\begin{center}
    \includegraphics[width=0.43\textwidth]{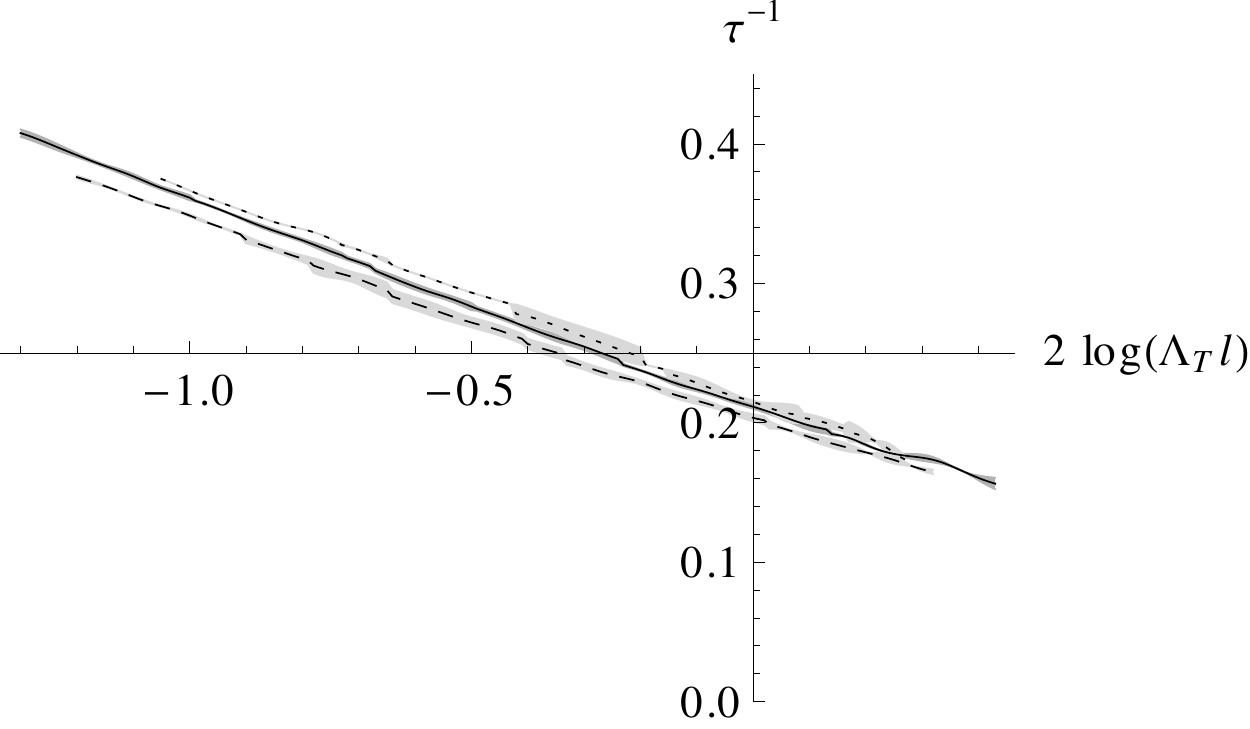}
\caption{Rapid convergence to $N=\infty$ for $\tau_N(l)$ in continuum. }
\label{fig-InvTau}
\end{center}
\end{figure}

The continuum functions $\frac 1{\tau_N(l)} $ are shown in Fig.~\ref{fig-InvTau}
for $N=11$, $19$, $29$ by a dashed line, a solid line, and a dotted line, respectively, 
and rapid convergence to $N=\infty$ is evident. 
The horizontal axis is labeled by $2\log(\Lambda_T l)$. $\Lambda_T$ 
is the infinite-$N$ critical
deconfinement temperature, roughly 
equivalent to 264 MeV in QCD units. The shades indicate the accumulated errors, 
for those regimes where we had enough data to reliably estimate them. By that we
mean that we had data points at least at three distinct pairs $(b,L)$ all corresponding to the
same physical $l$. The relation between $l$ and $(b,L)$ was set by a standard 
one-loop tadpole-improved formula, known to work well from other simulations. 
A reliable estimate of the continuum limit could be obtained by extrapolating in the
lattice spacing squared and observing a linear behavior. At $N=29$, only
two pairs were available for some values of $l$ and a continuum number was obtained by postulating a linear
behavior, but without an error. 
The approximate linearity of the continuum functions $\frac 1{\tau_N(l)}$ is consistent with
asymptotic freedom when $\tau_N(l)$ is viewed as an effective 
running 
coupling constant. The slope of the lines in Fig.~\ref{fig-InvTau} is about 0.22, while
the expected coefficient is about 0.29. The discrepancy
has to do with a factor associated with smearing, which 
has been so far only calculated at tree level where the theory
is conformal with a dependence only on the ratio $\frac {l^2}{s}$. 

Because of our indirect way to establish criticality, some of the
large-$N$ universality predictions we listed earlier become tautological. 
However, there remain two extra checks that are meaningful: one is to determine
the exponent $3/2$ from the ratio $a_0/a_1$ and the other is the exponent $3/4$
from the $N^{-3/4}$ level density in the critical region. Logarithmic 
fits produce estimates within 1\% of the expected values. 
This concludes our account of the numerical evidence for the non-analyticity.

Having this one example of a large-$N$ phase transition opens up
some new questions: Is the transition physical in the particle physics
sense, that is, would one actually see new singularities in a 
large-$N$, narrow-width approximation to the $S$-matrix of the theory?
If the answer is positive, one might speculate that at $N=\infty$ 
one does have exactly linear Regge trajectories for high spin states, up to a 
point, beyond which, presumably, a perturbative behavior ensues. If the answer is
negative, one would need to understand how exactly the $S$-matrix
gets shielded from the above large-$N$ phase transition we have established. 
In this context it is important to note that the non-analyticity we found does
not occur in the Wilson loop expectation values themselves, so
long as the number of boxes in the Young tableau of the representation 
is kept finite and fixed as $N\to\infty$. 

Many avenues for further investigations are likely to open up, providing 
some encouragement to those pursuing the long quest of conquering the large-$N$ limit of QCD, 
albeit, perhaps, in only a semi-analytic way.

We now turn to a review of the history of the
subject, focusing on the main contributions. 
We avoided doing this earlier in order to keep the presentation streamlined. 
The large-$N$ phase transition we discuss 
was discovered by Durhuus and Olesen~\cite{do} in the context
of two-dimensional $\SU(N)$ gauge theory, where Casimir dominance is exact. 
They obtained
the phase transition working directly at infinite $N$, where they 
identified the inviscid Burgers' equation as playing a central role. 
Therefore, it would be fair to refer to this transition as the DO transition.
Blaizot and Nowak~\cite{bn} argued that the large-$N$ universality class
was controlled by Burgers' equation~\cite{burgers}, including its viscous term.
This was shown to hold exactly in the \hk case~\cite{burgershk}. The
generic nature of the DO transition, in the sense that it occurs
also in three- and four-dimensional $\SU(N)$ gauge theory, was conjectured about
five years ago~\cite{conject}. A test in three dimensions was shown to be consistent
with the conjecture, but there was no overwhelming numerical evidence~\cite{threed}. Continuum
smearing was introduced at the same time, as a technical device used 
to define ``eigenvalues'' of Wilson loops in renormalized
QCD. The approximate validity of a ``diffusive'' viewpoint 
(that is, using the \hk as a model) for the behavior
of Wilson loops has been pointed out already in 2005~\cite{lenz}, 
for the case of the gauge 
group $\SU(2)$. The exact formula for the single eigenvalue-angle
density in the \hk case at any $N$ was derived in~\cite{tilo}. 
Casimir dominance has been discussed at the perturbative level
in~\cite{frenktaylor} and at the nonperturbative one it was reviewed 
by Greensite in~\cite{greensite}.
The limitation on the size of a finite box at infinite $N$ is studied 
in~\cite{contred}. 

The four-dimensional
test in this paper became possible after a hardware upgrade of
a computer cluster at Rutgers. Substantial resources were invested to
produce a convincing and detailed case in four dimensions, as the question
of whether a large-$N$ phase transition can occur in four-dimensional continuum 
$\SU(N)$ gauge theory, 
as the scale is varied, has been considered repeatedly in the past, 
but without definitive evidence one way or another~\cite{inst}.

The relation between the large-$N$ expansion and the program of dual 
topological unitarization~\cite{chewrosen} was explored a long time ago by Veneziano~\cite{gabrielle}. 
The associated question of Regge trajectory linearity has been
addressed by McGuigan and Thorn~\cite{thorn}. 
A more direct connection between the infinite-$N$ limit and 
zero string-coupling string theory has been an object of study for a
long time, and perhaps the best known line of attack is the suggestion to start from the
Makeenko-Migdal~\cite{mm} equations and guess a solution defined by a string theory.  

RL and HN acknowledge partial support by the DOE under grant
number DE-FG02-01ER41165. We are grateful to R.~Narayanan who was
involved in the early stages of this project.

\end{document}